\documentclass[prl,twocolumn,superscriptaddress,showpacs,floatfix]{revtex4}
\usepackage{graphicx,graphics,color,epsfig}
\usepackage{bm}
\usepackage{amsmath}
\usepackage{amssymb}
\usepackage{graphicx}

\begin{document}

\preprint{}
\title{Transport through the single-molecular dots in an external irradiation%
}
\author{Rong L\"{u}}
\affiliation{Center for Advanced Study, Tsinghua University, Beijing 100084, P. R. China}
\date{\today}

\begin{abstract}
We present a fully nonequilibrium calculation of the low-temperature
transport properties of a single molecular quantum dot coupled to local
phonon mode when an ac field is applied to the gate. The resonant behavior
is shown in the time-averaged differential conductance as the ac frequency
matches the frequency of the local phonon mode, which is a direct
consequence of the satellite-phonon-peak structure in the dot electron
spectral function. The different step structure with and without the
external irradiation is found in the I-V\ curves, and the oscillation
behavior is found in the step height as a function of the irradiation
intensity.
\end{abstract}

\maketitle

\bigskip Recent advances in nanotechnology have allowed the fabrication of
very small molecular quantum dots weakly coupled to the macroscopic charge
reservoirs (leads)\cite{molecule(e)}. In contrast to the semiconductor dot,
which is quite rigid in space, the molecules involved in the electron
tunneling process naturally possess the vibrational degrees of freedom which
will inevitably react to the transform of electrons through the molecular
quantum dots\cite{molecule(e)}. 
In addition to the importance in molecular-scale electronics from the application
point of view, these artificial, tunable devices are potentically important for
understanding the basic physics including the many-body effect.
Theoretically, a lot of effort has been
focused on the quantum conductance of molecular systems based on the kinetic
equation approach\cite{kinetic-eq}, the rate equation approach\cite{rate-eq}%
, the correlation effects\cite{correlation}, the nonequilibrium quantum
theory\cite{Wingreen89,Gogolin02,Zhu02,Lundin02}, and the strong coupling to
environment\cite{Flensberg02}. So far, the stationary quantum transport
through the molecular dots has been considered, while the influence of a
time-dependent ac field on the current has not been well addressed.
Irradiation of a quantum dot with an ac field\cite{Jauho94,Buettiker98}
offers a new way of affecting its dynamics, which enables one to study the
effect of electron-phonon interaction on the transport phenomenon of
molecular dots in essentially nonequilibrium condition.

In this Letter, we use the Keldysh nonequilibrium Green's function technique
to study the nonlinear ac transport through a single-molecular quantum dot
coupled to a local phonon mode with the external irradiation applied to the
gate, for the first time. After a canonical transformation, we obtain a
formula for the time-dependent current in general terms of bias,
temperature, the intensity and frequency of the external ac field, and the
electron-phonon coupling. We show that the satellite-peak structure due to
the electron-phonon interaction can be probed by imposing on top of gate
bias an ac bias voltage. The satellite-peak structure in the dot electron
spectral function gives rise to resonant behavior in the time-averaged
current as the ac frequency matches the frequency of local phonon mode,
which can be observed directly in experiments. The calculated I-V\ curves
also show the different step structures with and without the external
irradiation, and the step height shows the Bessel-type oscillation behavior
as a function of irradiation intensity.

In this work we consider a simplest Holstein-type model with a single phonon
mode is employed to address the vibrational degrees of freedom in the
molecular dot. All other complexity of real molecular devices, apart from
interaction with phonon mode, is ignored. Then the system Hamiltonian can be
written as%
\begin{equation}
H=H_{leads}+H_{X}+H_{D}+H_{T},
\end{equation}%
where%
\begin{eqnarray}
H_{leads} &=&\sum_{k,\eta ,\sigma }\epsilon _{k\eta }c_{k\eta \sigma
}^{\dagger }c_{k\eta \sigma },  \notag \\
H_{X} &=&\omega _{0}a^{\dag }a,  \notag \\
H_{D} &=&\sum_{\alpha }\left[ \epsilon _{0}\left( t\right) +\lambda \left(
a+a^{\dag }\right) \right] d_{\alpha }^{\dag }d_{\alpha },  \notag \\
H_{T} &=&\sum_{k,\eta ,\sigma ,\alpha }\left[ V_{k\sigma ,\alpha }^{\eta
}c_{k\eta \sigma }^{\dagger }d_{\alpha }+\text{H. C}.\right] .
\end{eqnarray}%
$c_{k\eta \sigma }^{\dagger }$ $\left( c_{k\eta \sigma }\right) $ are
creation (annihilation) operators for the noninteracting electrons with
momentum $k$ and spin index $\sigma $ in the left $\left( \eta =L\right) $
or right $\left( \eta =R\right) $ metallic leads. $\omega _{0}$ is the
frequency of the single phonon mode, and $a^{\dag }$ $\left( a\right) $ is
the phonon creation (annihilation) operator. $H_{D}$ describes the electron
in the quantum dot coupled to the local phonon mode with the coupling
constant $\lambda $, where $d_{\alpha }^{\dag }$ $\left( d_{\alpha }\right) $
is the dot-electron creation (annihilation) operator, and $\epsilon
_{0}\left( t\right) $ is the single energy level of the dot which can be
tuned by the external irradiation, $\epsilon _{0}\left( t\right) =\epsilon
_{0}+V_{r}\cos \left( \omega _{r}t\right) $ for harmonic bias. Here we
assume that the metallic leads are dc biased, neglecting the possible
\textquotedblleft leakage\textquotedblright\ of the irradiating ac field to
the leads. The generalization onto the case of nonzero ac bias is
straightforward. $H_{T}$ describes the tunneling coupling between the dot
and the leads, where the tunneling matrix elements $V_{k\sigma ,\alpha
}^{\eta }$ transfer electrons through an insulating barrier out of the dot.

Based on the Keldysh nonequilibrium Green's function formalism\cite{Keldysh65}%
, the time-dependent current from the $\eta $ lead to the dot is given by%
\cite{Jauho94}%
\begin{eqnarray}
J_{\eta }\left( t\right) &=&-\frac{2e}{\hbar }\sum_{\alpha }\int_{-\infty
}^{t}dt^{\prime }\int \frac{d\epsilon }{2\pi }\text{Im}\left\{ e^{i\epsilon
\left( t-t^{\prime }\right) }\Gamma _{\alpha }^{\eta }\right.  \notag \\
&&\left. \times \left[ G_{\alpha \alpha }^{<}\left( t,t^{\prime }\right)
+f_{\eta }\left( \epsilon \right) G_{\alpha \alpha }^{r}\left( t,t^{\prime
}\right) \right] \right\} ,  \label{current1}
\end{eqnarray}%
where $f_{L\left( R\right) }\left( \epsilon \right) $ are the Fermi
distribution function of the left (right) leads, which have different
chemical potentials upon a dc bias voltage $\mu _{L}-\mu _{R}=eV$. $\Gamma
_{\alpha }^{\eta }=2\pi \rho _{\alpha }\left( 0\right) \left\vert V_{k\alpha
,\alpha }^{\eta }\right\vert ^{2}$ characterizes the coupling between the
dot and the leads, and $\rho _{\alpha }\left( 0\right) $ is the spin-$\alpha 
$ band density of states in the leads. Here we assume that the leads give
rise to a flat, energy independent, density of states (i. e., the wide-band
limit). $G^{r\left( <\right) }$ is the retarded (lesser) Green's function of
the dot.

In order to compute the time-dependent current, one has to compute the dot
electron Green's functions in the presence of both the electron-phonon
interaction and the tunneling coupling between dot and leads. The Green's
function can be calculated by performing the canonical transformation $%
\mathcal{S=}\left( \lambda /\omega _{0}\right) \sum_{\alpha }d_{\alpha
}^{\dag }d_{\alpha }\left( a^{\dag }-a\right) $\cite{Mahan}, and then the
dot level is renormalized to $\epsilon _{0}-\Delta $, where $\Delta =\lambda
^{2}/\omega _{0}$, and the tunneling coupling term is also renormalized as $%
\overline{H}_{T}=\sum_{k,\eta ,\sigma ,\alpha }\left[ V_{k\sigma ,\alpha
}^{\eta }c_{k\eta \sigma }^{\dagger }d_{\alpha }X+H.C.\right] $, where $%
X=\exp \left[ -\left( \lambda /\omega _{0}\right) \left( a^{\dag }-a\right) %
\right] $. Ignoring the effects of narrowing the bands of leads due to the
phonons\cite{Hewson80}, the dot-electron Green's function can be decoupled
as $G_{\alpha \alpha ^{\prime }}^{r}\left( t,t^{\prime }\right) =\widetilde{G%
}_{\alpha \alpha ^{\prime }}^{r}\left( t,t^{\prime }\right) \left\langle
X\left( t\right) X^{\dag }\left( t^{\prime }\right) \right\rangle _{ph}$,
where $\widetilde{G}_{\alpha \alpha ^{\prime }}^{r}\left( t,t^{\prime
}\right) =-i\Theta \left( t-t^{\prime }\right) \left\langle \left\{ 
\widetilde{d}_{\alpha }\left( t\right) ,\widetilde{d}_{\alpha }^{\dag
}\left( t^{\prime }\right) \right\} \right\rangle _{el}$, $\widetilde{d}%
_{\alpha }\left( t\right) =e^{i\overline{H}_{el}t}d_{\alpha }e^{-i\overline{H%
}_{el}t}$, $\overline{H}_{el}=H_{X}$, and $X\left( t\right) =e^{i\overline{H}%
_{ph}t}Xe^{-i\overline{H}_{ph}t}$. The renormalization factor due to the
electron-phonon interaction is\cite{Mahan} $\left\langle X\left( t\right)
X^{\dag }\left( t^{\prime }\right) \right\rangle _{ph}=e^{-\Phi \left(
t-t^{\prime }\right) }$, where $\Phi \left( t\right) =\left( \lambda /\omega
_{0}\right) ^{2}\left[ N_{ph}\left( 1-e^{i\omega _{0}t}\right) +\left(
N_{ph}+1\right) \left( 1-e^{-i\omega _{0}t}\right) \right] $, and $N_{ph}=1/%
\left[ \exp \left( \beta \omega _{0}\right) -1\right] $. The retarded
Green's function can be easily obtained by the standard Dyson equation
approach\cite{Jauho94}, and the result is%
\begin{eqnarray}
G_{\alpha \alpha ^{\prime }}^{r}\left( t,t^{\prime }\right) &=&-i\delta
_{\alpha \alpha ^{\prime }}\Theta \left( t-t^{\prime }\right)
e^{-i\int_{t^{\prime }}^{t}d\tau V_{r}\cos \left( \omega _{r}\tau \right) } 
\notag \\
&&\times e^{-i\left[ \epsilon _{0}-\Delta -\frac{i}{2}\Gamma _{\alpha }%
\right] \left( t-t^{\prime }\right) -\Phi \left( t-t^{\prime }\right) },
\end{eqnarray}%
where $\Gamma _{\alpha }=\Gamma _{\alpha }^{L}+\Gamma _{\alpha }^{R}$ is the
total tunneling coupling to the leads. Following the operational rules\cite%
{Langreth76} to the Dyson equation for the contour-ordered Green's function,
the Keldysh Green's function is found to be%
\begin{eqnarray}
G_{\alpha \alpha ^{\prime }}^{<}\left( t,t^{\prime }\right) &=&\delta
_{\alpha \alpha ^{\prime }}\int dt_{1}\int dt_{2}\widetilde{G}_{\alpha
\alpha }^{r}\left( t,t_{1}\right)  \notag \\
&&\times \Sigma _{\alpha }^{<}\left( t_{1},t_{2}\right) \widetilde{G}%
_{\alpha \alpha }^{a}\left( t_{2},t^{\prime }\right) ,
\end{eqnarray}%
with the less self-energy%
\begin{equation}
\Sigma _{\alpha }^{<}\left( t_{1},t_{2}\right) =i\sum_{\eta }\int \frac{%
d\epsilon }{2\pi }\Gamma _{\alpha }^{\eta }e^{-\Phi \left[ -\left(
t_{1}-t_{2}\right) \right] }f_{\eta }\left( \epsilon \right) e^{-i\epsilon
\left( t_{1}-t_{2}\right) }.
\end{equation}%
Without electron-phonon interaction, the above result fully agrees with that
for time-dependent transport through noninteracting quantum dot\cite{Jauho94}%
.

Substitution of the Green's functions into Eq. (\ref{current1}) gives%
\begin{eqnarray}
&&J_{\eta }\left( t\right)  \notag \\
&=&-\frac{e}{\hbar }\sum_{\alpha }\Gamma _{\alpha }^{\eta }\int \frac{%
d\omega }{2\pi }\left\{ 2f_{\eta }\left( \omega \right) \text{Im}\left[
A_{\alpha }\left( \omega ,t\right) \right] \right.  \notag \\
&&-2\left( \sum_{\eta }\Gamma _{\alpha }^{\eta }f_{\eta }\left( \omega
\right) \right)  \notag \\
&&\left. \times \int_{-\infty }^{t}dt_{1}e^{-\Gamma _{\alpha }\left(
t-t_{1}\right) }\text{Im}\left[ A_{\alpha }\left( \omega ,t_{1}\right) %
\right] \right\} ,  \label{current2}
\end{eqnarray}%
with $A_{\alpha }\left( \omega ,t\right) =\int_{-\infty }^{t}dt^{\prime
}e^{i\omega \left( t-t^{\prime }\right) }G_{\alpha \alpha }^{r}\left(
t,t^{\prime }\right) $. Obviously, in the time-independent case, $A_{\alpha
}\left( \omega \right) $ is just the Fourier transform of the retarded
Green's function $G_{\alpha \alpha }^{r}\left( \omega \right) $. After some
algebra, we find that for this model,%
\begin{eqnarray}
&&A_{\alpha }\left( \omega ,t\right)  \notag \\
&=&e^{-g\left( 2N_{ph}+1\right) }\sum_{l,m,n}\left( -1\right)
^{l+m}e^{i\left( l-m\right) \omega _{r}t}e^{n\omega _{0}\beta /2}  \notag \\
&&\times J_{l}\left( \frac{V_{r}}{\omega _{r}}\right) J_{m}\left( \frac{V_{r}%
}{\omega _{r}}\right) I_{n}\left\{ 2g\left[ N_{ph}\left( N_{ph}+1\right) %
\right] ^{1/2}\right\}  \notag \\
&&\times \frac{1}{\omega -\left( \epsilon _{0}-\Delta \right) +m\omega
_{r}-n\omega _{0}+\frac{i}{2}\Gamma _{\alpha }},  \label{A}
\end{eqnarray}%
where $J_{m}\left( z\right) $ is the Bessel function of the $m$-th order, $%
I_{n}\left( z\right) $ is the Bessel function of complex argument, and $l$, $%
m$, $n=0$, $\pm 1$, $\pm 2\cdots $. Eq. (\ref{A}) together with the current
expression Eq. (\ref{current2}) provides the complete solution to the
time-dependent transport of molecular quantum dot coupled to local phonon
mode in the ac field. 
\begin{figure}[tbp]
\includegraphics[width=8.5cm]{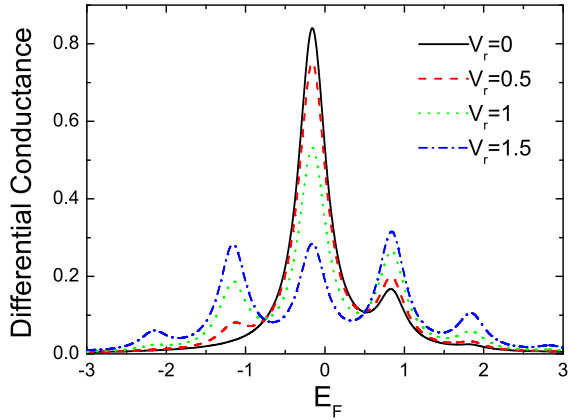}
\caption{The time-averaged differential conductance, in units of $2e^{2}/h$
through the molecular quantum dot as a function of the Fermi energy $E_{F}$
measured relative to the single level of the dot in the absence ($V_{r}=0$),
and in the presence of ac field with the frequency $\protect\omega _{r}=%
\protect\omega _{0}$ and three different values of irradiation intensities $%
V_{r}=0.5\protect\omega _{0}$, $\protect\omega _{0}$, and $1.5\protect\omega %
_{0}$, respectively. The energy is measured in units of the frequency of the
phonon mode $\protect\omega _{0}$, $\protect\lambda =0.4\protect\omega _{0}$%
, $\protect\epsilon _{0}=0$, and $\Gamma =0.2\protect\omega _{0}$.}
\end{figure}
Experimentally what is interest is the current on a time scale long compared to
$2\pi/\omega_{r}$. Here we discuss the time-averaged current $\left\langle J\left( t\right)
\right\rangle $, which
could be directly relevant to experiment. For this model, we then obtain%
\begin{eqnarray}
&&\left\langle J_{L}\left( t\right) \right\rangle =-\left\langle J_{R}\left(
t\right) \right\rangle  \notag \\
&=&\frac{2e}{\hbar }\left( \frac{1}{2}\right) e^{-g\left( 2N_{ph}+1\right)
}\sum_{\alpha }\Gamma _{\alpha }^{L}\Gamma _{\alpha }^{R}\int \frac{d\omega 
}{2\pi }  \notag \\
&&\times \left[ f_{L}\left( \omega \right) -f_{R}\left( \omega \right) %
\right] \sum_{m,n}J_{m}^{2}\left( \frac{V_{r}}{\omega _{r}}\right)  \notag \\
&&\times I_{n}\left( 2g\left[ N_{ph}\left( N_{ph}+1\right) \right]
^{1/2}\right) e^{n\omega _{0}\beta /2}  \notag \\
&&\times \frac{1}{\left[ \omega -\left( \epsilon _{0}-\Delta \right)
+m\omega _{r}-n\omega _{0}\right] ^{2}+\left( \frac{\Gamma _{\alpha }}{2}%
\right) ^{2}}.  \label{t-averaged-current}
\end{eqnarray}%
In the absence of external irradiation, i.e., $V_{r}=0$, Eq. (\ref%
{t-averaged-current}) fully agrees with the result of dc current in Refs. 
\cite{Zhu02,Lundin02}. The contribution of the external irradiation to the
transport becomes significant when the argument of the Bessel function, $%
V_{r}/\omega _{r}$, is of the order of unity. Another important consequence
is that the time-averaged current under irradiation is proportional to $%
J_{m}^{2}$. As a result, one should expect the change in current due to the
external irradiation to depend on the intensity of irradiation. 
\begin{figure}[tbp]
\includegraphics[width=8.5cm]{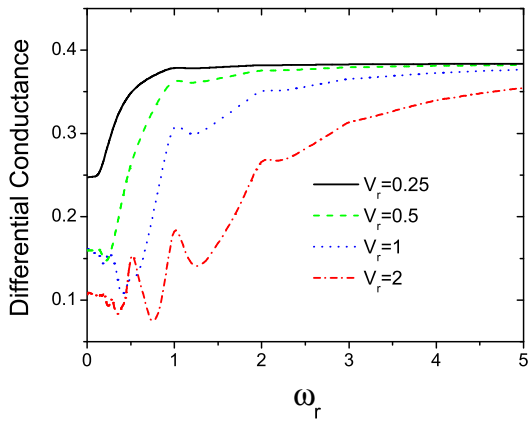}
\caption{The time-averaged differential conductance, in units of $2e^{2}/h$
as a function of the frequency of irradiating ac field for different
intensities of ac field with $V_{r}/\protect\omega _{0}=0.25$, $0.5$, $1$,
and $2$, respectively. Here $\protect\lambda /\protect\omega _{0}=1$, $%
\protect\epsilon _{0}=0$, $\Gamma =0.2\protect\omega _{0}$, and $%
E_{F}=\left( -1\right) \protect\omega _{0}$.}
\end{figure}

For simplicity, we consider the tunneling coupling between the molecular dot
and the two leads to be symmetric and independent of the spin index, i.e., $%
\Gamma _{\uparrow }^{L}=\Gamma _{\downarrow }^{L}=\Gamma _{\uparrow
}^{R}=\Gamma _{\downarrow }^{R}=\Gamma $. In Fig. 1, we plot the
zero-temperature differential conductance as a function of the Fermi energy
measured relative to the single level $\epsilon _{0}$ of the dot in the
external ac field with frequency $\omega _{r}=\omega _{0}$ and different
values of irradiation intensities. For comparison, we also plot the
differential conductance in the absence of ac field. At zero ac bias
voltage, our results agrees well with that in Refs. \cite{Zhu02,Lundin02},
where the electron-phonon coupling can lead to the satellite resonant peaks. 
Fig. 1 shows that the ac field with frequency $\omega
_{r}=\omega _{0}$ can lead to the enhancement of satellite resonant
peaks in the positive energy region and the appearance of new
peaks in the negative energy region. One can also see that the main peak is
suppressed by the irradiation while increasing the intensity. In Fig. 2, we
plot the zero-temperature differential conductance as a function of the
frequency of ac field for different values of irradiation intensities by
fixing the Fermi energy of the leads $E_{F}$ as $E_{F}-\left( \epsilon
_{0}-\Delta \right) =0$ to avoid unnecessary complications. As the intensity
of ac field increases, resonant signals are clearly shown when the frequency
of ac field satisfies $m\omega _{r}=n\omega _{0}$. The satellite-phonon-peak
structure in the dot electron spectral function, as shown in Refs. \cite%
{Zhu02,Lundin02} and also in Fig. 1 of this Letter, gives rise to resonant
behavior in the conductance as the irradiating ac frequency matches the
frequency of local phonon mode, and can be observed directly in experiments.
Fig. 2 also shows that more resonant signals can be observed while
increasing the irradiation intensity. 
\begin{figure}[tbp]
\includegraphics[width=8.5cm]{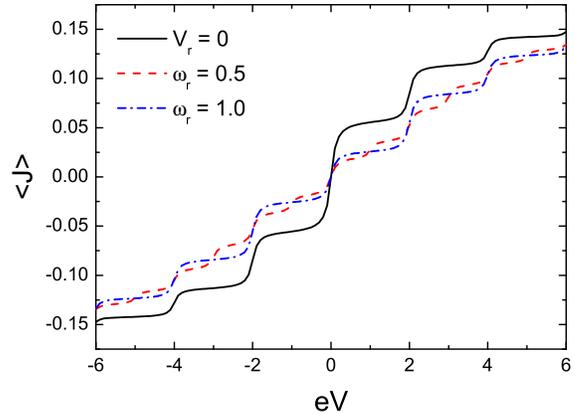}
\caption{Current-voltage characteristics, in units of $2e/h$, without ($%
V_{r}=0$) and with the ac field for the different irradiative frequency: $%
\protect\omega _{r}=0.5\protect\omega _{0}$ and $1\protect\omega _{0}$.
Here, $\protect\lambda /\protect\omega _{0}=1$, $E_{F}/\protect\omega _{0}=-1
$, $\protect\epsilon _{0}=0$, $V_{r}/\protect\omega _{0}=2$, and $\Gamma
=0.05\protect\omega _{0}$.}
\end{figure}
\begin{figure}[tbp]
\includegraphics[width=8.5cm]{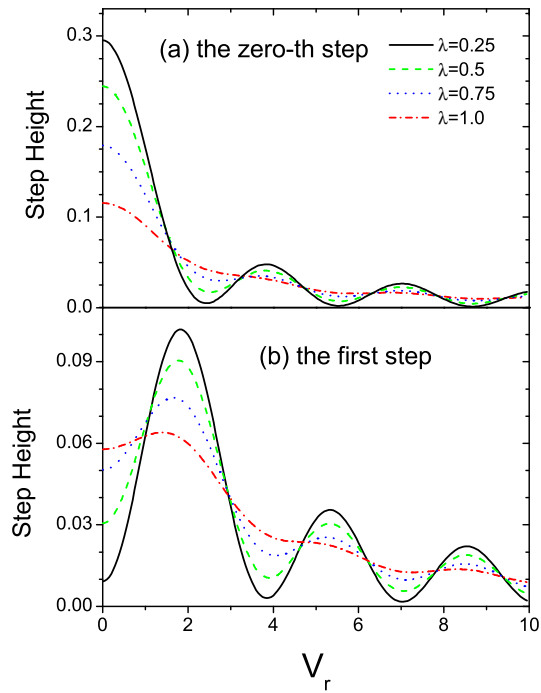}
\caption{The height of the zeroth ($N=0$) [a] and the first ($N=1$) [b]
step, in units of $2e/h$, as a function of the irradiation intensity by
fixing the Fermi energy of the leads $E_{F}$ as $E_{F}-\left( \protect%
\epsilon _{0}-\Delta \right) =0$. Here, $\protect\omega _{r}=\protect\omega %
_{0}$, and $\Gamma =0.05\protect\omega _{0}$.}
\end{figure}

Fig. 3 shows the calculated zero-temperature current-voltage curves with and
without the external ac irradiation. Here we assume the leads to be
symmetrically voltage biased, i.e., $+V/2$ on the left lead and $-V/2$ on
the right one, to avoid unnecessary complications. In the absence of
external ac field, clear steps appear at roughly $2\omega _{0}$ intervals in
the weak tunneling coupling limit, corresponding to $eV/2=\pm n\omega _{0}$,
with $n=0$, $1$, $2\cdots $, and the height of the $N$-th step decreases
with $N$, which can be easily understood from Eq. (\ref{height}) [see below]
by taking $V_{r}=0$ for small electron-phonon interaction. In the presence
of ac field with frequency $\omega _{r}=\omega _{0}$, steps appear at the
same intervals $2\omega _{0}$ as those in the zero ac field, while the step
height is modulated by Bessel function due to the irradiation [see Eq. (\ref%
{height}) in below]. Fig. 3 also shows that more steps appear at the
intervals $\omega _{0}$ in the case of $\omega _{r}=0.5\omega _{0}$,
corresponding to $eV/2=m\omega _{r}-n\omega _{0}$ with $m=0$, $\pm 1$, $\pm
2\cdots $. The zero-temperature $N$-th step's height in an external ac field
with frequency $\omega _{r}=\omega _{0}$ can be analytically obtained from
Eq. (\ref{t-averaged-current}) at the fixed Fermi energy $E_{F}$ as $%
E_{F}-\left( \epsilon _{0}-\Delta \right) =0$,%
\begin{eqnarray}
\Delta J_{N} &=&\pi \left( \frac{2e}{h}\right) e^{-g}\sum_{\alpha }\frac{%
\Gamma _{\alpha }^{L}\Gamma _{\alpha }^{R}}{\Gamma _{\alpha }^{L}+\Gamma
_{\alpha }^{R}}  \notag \\
&&\times \sum_{n=0}^{\infty }\frac{g^{n}}{n!}\left[ J_{n+N}^{2}\left( \frac{%
V_{r}}{\omega _{r}}\right) +J_{n-N}^{2}\left( \frac{V_{r}}{\omega _{r}}%
\right) \right] .  \label{height}
\end{eqnarray}%
In Fig. 4, we plot the height of the $N$-th step, where $N=0$ and $1$, as a
function of the irradiation intensity for different values of the
electron-phonon coupling constant. The oscillation behavior of the step
height is clearly observed for the small electron-phonon interaction due to
the external irradiation [see Eq. (\ref{height})], while this oscillation
smears out for large electron-phonon interaction. The large electron-phonon
interaction enhances the processes of absorption and emission of many $%
\left( n>1\right) $ phonons when electron tunnels through the dot, and then
the summation of Bessel function with large values of indices $n\pm N$
results in the smearing of the oscillation behavior [see Eq. (\ref{height})].

Summarizing, using the Keldysh nonequilibrium Green's function technique, we
have studied in this work, to the best of our knowledge, for the first time
the time-dependent transport through a single molecular quantum dot coupled
to a local phonon mode in the presence of an external ac field. We show that
the external irradiation provides another important experimental tool where
both the equilibrium and out of equilibrium transport phenomenon can be
probed. In particular, resonant behavior as the ac frequency matches the
frequency of local phonon mode is shown to exist as a result of the
satellite-phonon-peak structure in the dot electron density of states. The
nonlinear I-V curves exhibit new structure caused by the external
irradiation, which can be investigated experimentally.

We thank J.-X. Zhu for stimulating discussions which lead to this work and
Z.-R. Liu for useful discussions.

\end{document}